# Inconsistency of Carnot's theorem's proof by R. Clausius

V. Ihnatovych

Department of Philosophy, National Technical University of Ukraine "Kyiv Polytechnic Institute",

Kyiv, Ukraine

e-mail: V.Ihnatovych@kpi.ua

## Abstract

R. Clausius proved Carnot's theorem basing on postulate: «Heat cannot, of itself, pass from a colder to a hotter body». Alexander Gukhman demonstrated that Carnot's theorem can be proved based on the postulate: «Heat cannot, of itself, pass from a hotter to a colder body». He concluded that Carnot's theorem does not follow from Clausius's postulate. The following paper gives a detailed justification of Gukhman's derivation.

## 1. Introduction

In his work «Reflections on the motive power of heat and on machines fitted to develop this power» [1] Sadi Carnot proves the theorem, according to which maximum work made by heat engine does not depend on the properties of working body used in the engine. He wrote «*The motive power of heat is independent of the agents employed to realize it; its quantity is fixed solely by the temperatures of the bodies between which is effected, filially, the transfer of the caloric*» [1, p.68]. Carnot based the proof of this theorem on the statement of existence of indestructible caloric and postulate of the impossibility of getting work out of nothing. [1].

R. Clausius rejected the existence of caloric, and gave a new proof of Carnot's theorem. This evidence is well known, is included in many courses of thermodynamics and was not doubted by anyone for a long time.

A.A.Gukhman proved that Clausius's proof is incorrect. [3, 4]. His works were published in 1947 and 1986, but they are still not well-known nowadays. The author would like to draw the attention of researchers to the important results of the research of A.A.Gukhman.

The detailed proof of Carnot's theorem by Clausius is given below, and then the inconsistency of this evidence is shown in the same manner as in the works of Gukhman but in more details.

## 2. Derivation of Carnot's theorem based on Clausius's postulate

Clausius puts the proof of Carnot's theorem in a chapter «Second main Principle of the Mechanical Theory of Heat» of his work «The Mechanical Theory of Heat» [2].

The first paragraph of this chapter is called «Description of a special form of Cyclical Process». Here the Carnot's cycle is described. It's a cycle process that includes the following four processes: 1) isothermal enlargement of changing (working) body at temperature $T_1$, which equals to the temperature of reservoir of heat $K_1$; 2) adiabatic enlargement of changing body, in the process of which it's temperature lowers to the temperature $T_2$, which equals to the temperature of reservoir of heat $K_2$; 3) isothermal compression of working body at temperature $T_2$; 4) adiabatic compression of changing body, in the process of which it's temperature lowers to the temperature $T_1$.

After the end of the cyclic process changing body comes back to the initial state.

Second paragraph is called «*Result of the Cyclical Process*». There Clausius writes that in the process of first enlargement (at constant temperature $T_1$) changing body got $Q_1$ amount of heat from body $K_1$. In the process of first compression (at constant temperature $T_2$) changing body gives off $Q_2$ amount of heat from body $K_2$. During the cycle work $W$ is done, and $Q$ amount of heat is spent on it.

The result of cyclical process is described by Clausius as follows:

«*The one quantity of heat Q, derived from the body $K_1$, is transformed into work, and the other quantity $Q_2$ has passed over from the hotter body $K_1$ into the colder $K_2$*» [2, p.72].

He also writes the equation:

$Q_1 = Q_2 + Q$.

Further Clausius writes that described cyclical process can flow in the opposite direction. Clausius writes the following:

«In addition the variable body has drawn the quantity of heat $Q_2$ from the body $K_2$, and has given out to the body $K_1$ the quantity of heat $Q_1 = Q_2 + Q$. Of the two parts of which $Q_1$ consists, the one $Q$ corresponds to the work absorbed, and is generated from it, whilst the other $Q_2$ has passed over as heat from the body $K_2$ to the body $K_1$. Hence the result of the cyclical process may here be described as follows: the quantity of heat $Q$ is generated out of work, and is given off to the body $K_1$ and the quantity of heat $Q_2$ has passed over from the colder body $K_2$ to the hotter body $K_1$» [2, p.73].

In paragraph three «*Cyclical process in the case of a body composed partly of liquid, partly of vapour*» Clausius asks a question: «*Whether the quantity of heat converted into work, or generated out of work, stands in a generally constant proportion to the quantity which passes over from the hotter to the colder body, or vice versa; or whether the proportion existing between them varies according to the nature of the variable body, which is the medium of the transfer*» [2, p.76].

In paragraph five «*New Fundamental Principle concerning Heat*» Clausius formulates and explains principle: «Heat cannot, of itself, pass from a colder to a hotter body» [2, p.78]; in different words: «A passage of heat from a colder to a hotter body cannot take place without compensation» [2, p.78].

After that goes paragraph six called «*Proof that the relation between the quantity of heat carried over, and that converted into work, is independent of the nature of the matter which forms the medium of the change*».
Then the proof of this statement follows. The whole proof is given below with some words and symbols highlighted.

«Let there, if possible, be two bodies $C$ and $C'$ (e.g. the perfect gas and the combined mass of liquid and vapour, described above) for which the values of $Q$ are equal, but those of the transferred quantities of heat are different, and let these different values be called $Q_2$, and $Q'_2$ respectively: $Q'_2$ being the **greater** of the two. Now let us in the first place subject the body $C$ to a cyclical process, such that the quantity of heat $Q$ is transformed into work, and the quantity $Q_2$ is transferred from $K_1$ to $K_2$. Next let us subject $C'$ to a cyclical process of the reverse description, so that the quantity of heat $Q$ is generated out of work, and the quantity $Q'_2$ is transferred from $K_2$ to $K_1$. Then the above two changes, from heat into work, and work into heat, will cancel each other; since we may suppose that when in the first process the heat $Q$ has been taken from the body $K_1$ and transformed into work, this same work is expended in the second process in producing the heat $Q$, which is then returned to the same body $K_1$. In all other respects also the bodies will have returned, at the end of the two operations, to their original condition, with one exception only. The quantity of heat $Q'_2$, transferred from $K_2$ to $K_1$, has been assumed to be **greater** than the quantity $Q_2$ transferred from $K_1$ to $K_2$. Hence these two do not cancel each other, but there remains at the end a quantity of heat, represented by the difference $\boldsymbol{Q'_2 - Q_2}$, which has passed over from $\boldsymbol{K_2}$ to $\boldsymbol{K_1}$. Hence a passage of heat will have taken place from a **colder** to a **warmer** body without any other compensating change. But this contradicts the fundamental principle. Hence the assumption that $Q'_2$ is **greater** than $Q_2$ must be false.

Again, if we make the opposite assumption, that $Q'_2$ is **less** than $Q_2$, we may suppose the body $C'$ to undergo the cyclical process in the first, and $C$ in the reverse direction. We then arrive similarly at the result that a quantity of heat $\boldsymbol{Q_2 - Q'_2}$ has passed from the **colder** body $K_2$ to the **hotter** $K_1$, which is again contrary to the principle.

Since then $Q'_2$ can be neither greater nor less than $Q_2$ it must be equal to $Q_2$; which was to be proved» [2, p.80].

## 3. The proof of Carnot's theorem based on postulate opposite to Clausius's postulate

A. A. Gukhman showed that it's possible to prove that relation $Q/Q_2$ is independent from the kind of changing body if using "antipostulate": «Heat cannot, of itself, pass from a hotter to a colder body» [3, p.80] (see also [4, p.340]).

He wrote: «Replacement of Clausius's postulate (initial statement) with its antithesis (physically absurd thesis contrary to premise) is reflected neither on the results nor on the way they were obtained [4, p.341].

However A. A. Gukhman justified this statement briefly. Perhaps for this reason, the result went unnoticed.

To convincingly demonstrate the feasibility of proof based on the Carnot «antipostulate» «Heat cannot, of itself, pass from a hotter to a colder body», we repeat the above Clausius's argument, replacing the highlighted words with the opposite, and interchanging the highlighted characters.

«Let there, if possible, be two bodies $C$ and $C'$ (e.g. the perfect gas and the combined mass of liquid and vapour, described above) for which the values of $Q$ are equal, but those of the transferred quantities of heat are different, and let these different values be called $Q_2$, and $Q_2'$ respectively: $Q_2'$ being the **less** of the two. Now let us in the first place subject the body $C$ to a cyclical process, such that the quantity of heat $Q$ is transformed into work, and the quantity $Q_2$ is transferred from $K_1$ to $K_2$. Next let us subject $C'$ to a cyclical process of the reverse description, so that the quantity of heat $Q$ is generated out of work, and the quantity $Q_2'$ is transferred from $K_1$ to $K_2$. Then the above two changes, from heat into work, and work into heat, will cancel each other; since we may suppose that when in the first process the heat $Q$ has been taken from the body $K_1$ and transformed into work, this same work is expended in the second process in producing the heat $Q$, which is then returned to the same body $K_1$. In all other respects also the bodies will have returned, at the end of the two operations, to their original condition, with one exception only. The quantity of heat $Q_2'$, transferred from $K_2$ to $K_1$, has been assumed to be **less** than the quantity $Q_2$ transferred from $K_1$ to $K_2$. Hence these two do not cancel each other, but there remains at the end a quantity of heat, represented by the difference $\boldsymbol{Q_2' - Q_2}$ which has passed over from $\boldsymbol{K_1}$ to $\boldsymbol{K_2}$. Hence a passage of heat will have taken place from a **warmer** to a **colder** body without any other compensating change. But this contradicts the fundamental principle. Hence the assumption that $Q_2'$ is **less** than $Q_2$ must be false.

Again, if we make the opposite assumption, that $Q_2'$ is **greater** than $Q_2$, we may suppose the body $C'$ to undergo the cyclical process in the first, and $C$ in the reverse direction. We then arrive similarly at the result that a quantity of heat $\boldsymbol{Q_2' - Q_2}$ has passed from the **hotter** body $K_2$ to the **colder** $K_1$, which is again contrary to the principle.

Since then $Q_2'$ can be neither greater nor less than $Q_2$ it must be equal to $Q_2$; which was to be proved».

## 4. Discussion

A. A. Gukhman made a conclusion: «A thought arises that final conclusions do not logically depend on logical initial premises» [3, p.79; 4, p.341].

I think he was too cautious with his conclusion. Two opposite judgments cannot be followed by the same true conclusion.

A. A. Gukhman convincingly demonstrated that Carnot's theorem does not follow from Clausius's postulate.

## 5. Conclusions

The proof of Carnot's theorem made by Clausius is incorrect. Carnot's theorem does not follow from Clausius's postulate «Heat cannot, of itself, pass from a colder to a hotter body». Clausius's proof should be taken out from thermodynamics courses.

## 6. References

1. Carnot N.-L.-S. Reflections on the motive power of heat and on machines fitted to develop this power / Edited by R. H. Thurston, New York: John Wiley & Sons. London: Chapman & Hall, Limited. 1897.

2. Clausius R. The Mechanical Theory of Heat. London. Macmillan and Co. 1879.

3. Gukhman A. A. Ob osnovaniyakh termodinamiki (On the grounds of thermodynamics), Alma-Ata, Kazakh Academy of Sciences Publishers, 1947. [Russian]

4. Gukhman A. A. Ob osnovaniyakh termodinamiki (On the grounds of thermodynamics). – Moscow, Energoatomizdat Publishers, 1986. [Russian]